\documentclass[pra,twocolumn,showpacs,preprintnumbers,amsmath,amssymb]{revtex4-1}
\makeatletter
\usepackage[dvips]{graphicx}

\usepackage{amsmath}
\usepackage{amsbsy}
\usepackage{color}
\setcitestyle{square}
\usepackage{epstopdf}
\DeclareGraphicsExtensions{.pdf,.eps,.png,.jpg,.mps}

\definecolor{textcolor}{cmyk}{0,0,0,1}
\definecolor{magenta}{rgb}{1,0,1}
\definecolor{green}{rgb}{0,1,0}
\definecolor{red}{rgb}{1,0,0}

\begin{document}

\title{
Existence of nontrivial topologically protected states at grain boundaries in bilayer graphene: signatures and electrical switching  }
\author{W. Jask\'olski}
\affiliation{Institute of Physics, Faculty of Physics, Astronomy and Informatics, Nicolaus Copernicus University, Grudziadzka 5, 87-100 Toru\'n, Poland}
\author{M. Pelc}
\affiliation{Institute of Physics, Faculty of Physics, Astronomy and Informatics, Nicolaus Copernicus University, Grudziadzka 5, 87-100 Toru\'n, Poland}
\affiliation{Centro de F\'isica de Materiales, CFM-MPC CSIC-UPV/EHU, Donostia International Physics Center (DIPC)}
\author{Leonor Chico}
\affiliation{Instituto de Ciencia de Materiales de Madrid (ICMM-CSIC), Consejo Superior de Investigaciones Cient\'ificas, C/ Sor Juana In\'es de la Cruz 3, 28049 Madrid, Spain}
\author{A. Ayuela}
\affiliation{Centro de F\'isica de Materiales, CFM-MPC CSIC-UPV/EHU, Donostia International Physics Center (DIPC)}
\affiliation{Departamento de F\'isica de Materiales, Facultad de Qu\'imicas, UPV-EHU, 20018 San Sebasti\'an, Spain}

\date{\today}

%%%%%%%%%%%%%%%%%%%%%%%%%%%%%%%%%%%%%%%%%%%%%%%%%%%%%%%%%%%%%%%%%%%%%%%%%%%%%%%%
\begin{abstract}
Recent experiments [L. Ju et al., Nature, 2015, 520, 650] confirm the existence of gapless states at domain walls created in gated bilayer graphene, when the sublattice stacking is changed from AB to BA. These states are significant because they are topologically protected, valley-polarized and give rise to conductance along the domain wall. Current theoretical models predict the appearance of such states only at domain walls, which preserve the sublattice order. Here we show that the appearance of the topologically protected states in stacking domain walls can be much more common in bilayer graphene, since they can also emerge in unexpected geometries, e.g., at grain boundaries with atomic-scale topological defects.  We focus on a bilayer system in which one of the layers contains a line of octagon-double pentagon defects, that mix graphene sublattices.  We demonstrate that  gap states are preserved even with pentagonal defects. Remarkably, unlike previous predictions, the number of gap states changes by inverting the gate polarization, yielding an asymmetric conductance along the grain boundary under gate reversal. This effect, linked to defect states, should be detectable in transport measurements and could be exploited in electrical switches.\end{abstract}


\maketitle

%%%%%%%%%%%%%%%%%%%%%%%%%%%%%%%%%%%%%%%%%%%%%%%%%%%%%%%%%%%%%%%%%%%%%%%%%%%%%%%%
%%% INTRODUCTION %%%
%%%%%%%%%%%%%%%%%%%%%%%%%%%%%%%%%%%%%%%%%%%%%%%%%%%%%%%%%%%%%%%%%%%%%%%%%%%%%%%%

\section{Introduction}
Although graphene presents a record-high electron mobility at room temperature, its application in next-generation electronics requires the opening of an energy gap \cite{Bolotin_2008}. This has triggered the quest for various physical mechanisms capable of opening a substantial gap, including external fields, quantum size effects and the use of different substrates. Gated Bernal-stacked bilayer graphene (BLG) shows a tunable band gap \cite{Ohta_2006,Oostinga_2008,Zhang_2009}.  This setup is not only an interesting candidate for novel electronic applications \cite{fet, Lin_2008, Choi_2010, Padilha_2011} but also for the study of fundamental physical phenomena\cite{Maher_2013,Martin_2008,Zhang_2013,Vaezi_2013,bec}. The gate-induced gap in bilayer graphene has a topological origin \cite{Martin_2008}, linked to the existence of two inequivalent valley---$K$ and $K'$---which give rise to chiral edge and boundary states. Provided there is no-valley mixing, gated bilayer graphene presents topologically protected gapless states when a domain wall is introduced, either by an electric field reversal \cite{Martin_2008} or by an AB/BA stacking change \cite{Lin_2013,Alden_2013,Ju_2015, Pelc_2015}.  Such states are of high interest because they open a possibility of exploring valley physics in graphene.

The existence of stacking boundaries in BLG has been experimentally identified \cite{Lin_2013,Alden_2013}.  Since the separation in momentum space between graphene valleys is large, topologically protected, valley-polarized gap states may appear and localize at such domain walls; this fact has been also confirmed \cite{Ju_2015}. These states have been predicted so far at stacking boundaries created when one layer is either stretched or corrugated \cite{Lin_2013,Alden_2013,San_Jose_2014, Ju_2015}. In both cases the stacking change between graphene layers is introduced keeping its hexagonal structure, without introducing topological defects. This implies that sublattice symmetry is preserved. In these proposals, atomic-scale defects are avoided because, intuitively, they may act as sources of inter-valley scattering, which in principle can destroy topological gapless states \cite{Ju_2015}.

%% FIGURE %%%
\begin{figure}[hb]
\centering
\includegraphics[width=\columnwidth]{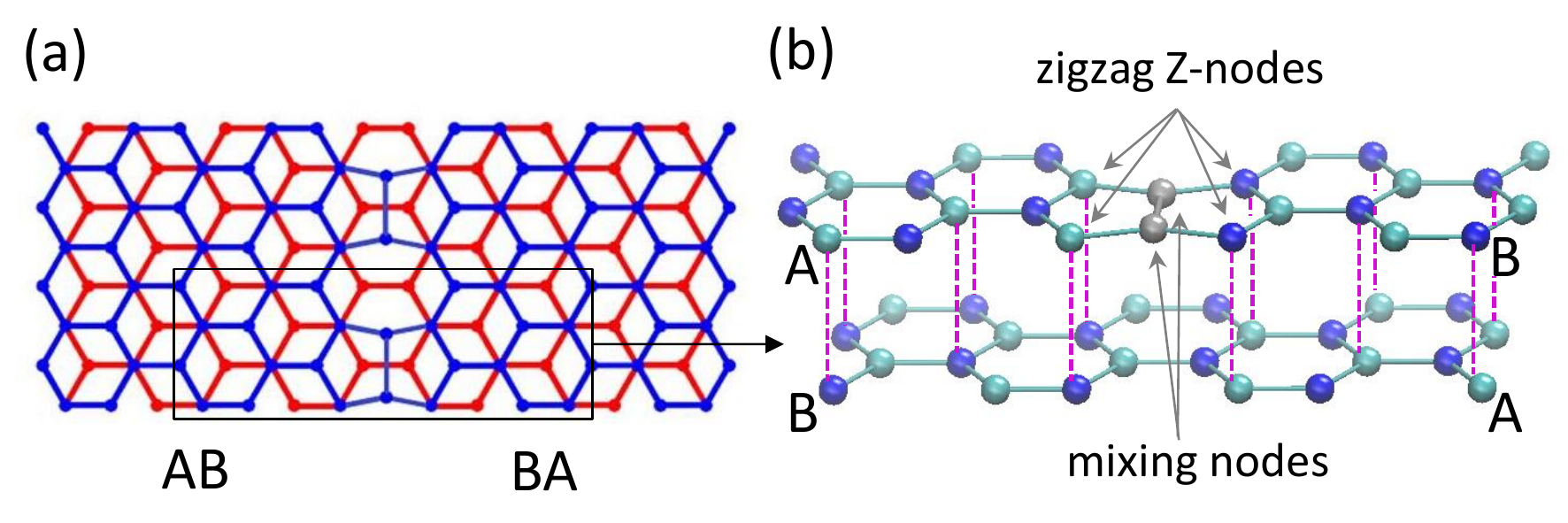}
\caption{\label{fig:fig1}
(Color online). (a) Top view of bilayer graphene with an AB/BA stacking boundary produced by an octagonal-pentagonal defect line in the top layer (blue). (b) Detailed view of the AB-BA stacking. Different sublattices are colored in blue and cyan. The nodes which mix the two sublattices are depicted in grey color. Neighboring nodes at the defect line in the top layer, called zigzag or Z-nodes, are also marked with arrows.
}
 \end{figure}

In this work we show that this intuition fails. We demonstrate that robust gap states can exist in BLG despite the presence of atomic-scale defects. Remarkably, some of these gap states are topologically protected states arising at stacking boundaries. We investigate a gated bilayer system, in which the top layer contains a defect line along a zigzag direction, built of octagons and pentagon pairs, and the bottom layer is pristine graphene. This type of defect line has been experimentally identified in monolayer graphene \cite{Lahiri_2010}. Our calculations reveal the existence of bands in the bulk gap, the number of which depends on the gate polarization. One of these bands is topologically protected, connecting the bulk valence and conduction bilayer continua. The other states stem from the mixture of the band induced by the defect line with those originated by the stacking change. Remarkably, despite the presence of pentagonal defects, which mix graphene sublattices, topologically protected gap states persist in this system, although sublattice symmetry is not conserved. This means that topological phases in BLG can be found also at grain boundaries, where they are not expected to occur.

\section{Model and methods}

The model system is schematically presented in Fig. \ref{fig:fig1}. On the two sides of the grain boundary, we have bilayer graphene with the most stable Bernal stacking, i.e., the nodes of one sublattice in one layer fall into hexagon centers of the other layer ({\it unconnected} sublattices), so the atoms of the other sublattices are vertically aligned ({\it connected} sublattices). 

Away from the defect line, we identify the stackings as AB and BA, as seen in Fig. \ref{fig:fig1}. The two semiinfinite graphenes in the top layer are joined by a line of octagons and double pentagon defects (8-55), forming the grain boundary. The pentagonal rings mix graphene sublattices. If one tries to label the two sublattices of the top layer from the outer parts towards the defect line, one ends up with two atoms in the unit cell which cannot be uniquely assigned to any sublattice, so the system presents geometrical frustration. These two bonded atoms, at the sides shared by the adjacent pentagons, are responsible for the sublattice mixing; we denominate them {\it mixing} nodes, see Fig. \ref{fig:fig1} (b). The nodes closest to the mixing ones are called zigzag or Z-nodes. This name derives from one possible way of creating an 8-55 boundary, namely by joining a zigzag-terminated edge \cite{our_prb_2013}. Note that the mixing nodes belong to the unconnected sublattice, whereas the Z-nodes belong to the connected one. 

 We perform calculations in one-orbital tight-binding approximation  for nearest-neighbors, with intralayer hopping parameter $\gamma_{0}= -2.66$ eV, and the interlayer hopping parameter, given only for nodes that are directly on top of each other, is set to $\gamma_{1}=0.1\gamma_{0}$ \cite{Castro_2007,Ohta_2006}. Note that the effects of screening, which mainly lead to a reduction of the energy gap, can be just seen as a scaling of $V$, but do not affect the global features of the gap states under study \cite{Castro_2007}. A gate voltage $V=\pm 0.3$ V is applied to the bottom layer. Based on this model, we calculate the local density of states (LDOS) \cite{misc:LDOS} using the Green function matching approach  \cite{Datta,Chico_1996,Jaskolski_2005}. As the system has translational symmetry along the grain boundary, the LDOS  is $k$-dependent in this direction. We have performed test calculations for different values of $V$ and $\gamma_1$, applying also the gate voltage to the top layer, and we have verified that the main results presented in Sec. \ref{sec:res}  do not depend qualitatively on these parameters.  

%% FIGURE %%%
\begin{figure}[thpb]
\centering
\includegraphics[width=\columnwidth]{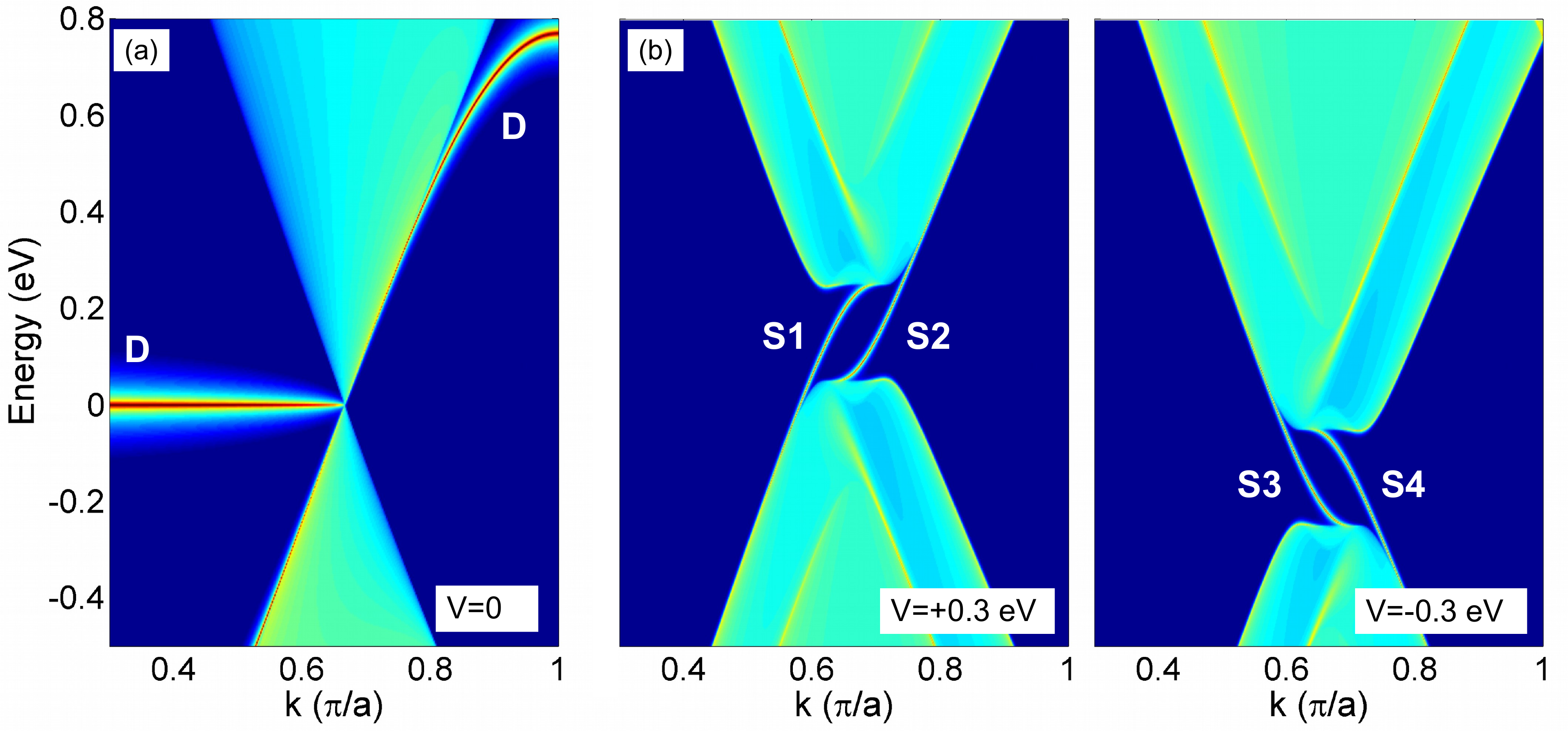}
\caption{\label{fig:ingr}
(Color online). 
(a) LDOS($E,k$) for monolayer graphene with an 8-55 grain boundary. (b)  LDOS($E,k$) for AB/BA stacked graphene with a minimal stacking boundary for positive and negative gate voltages. Topological gap states reverse the sign of their velocities when the voltage sign is changed. 
}
 \end{figure}

%% FIGURE %%%
\begin{figure}[thpb]
\centering
\includegraphics[width=\columnwidth]{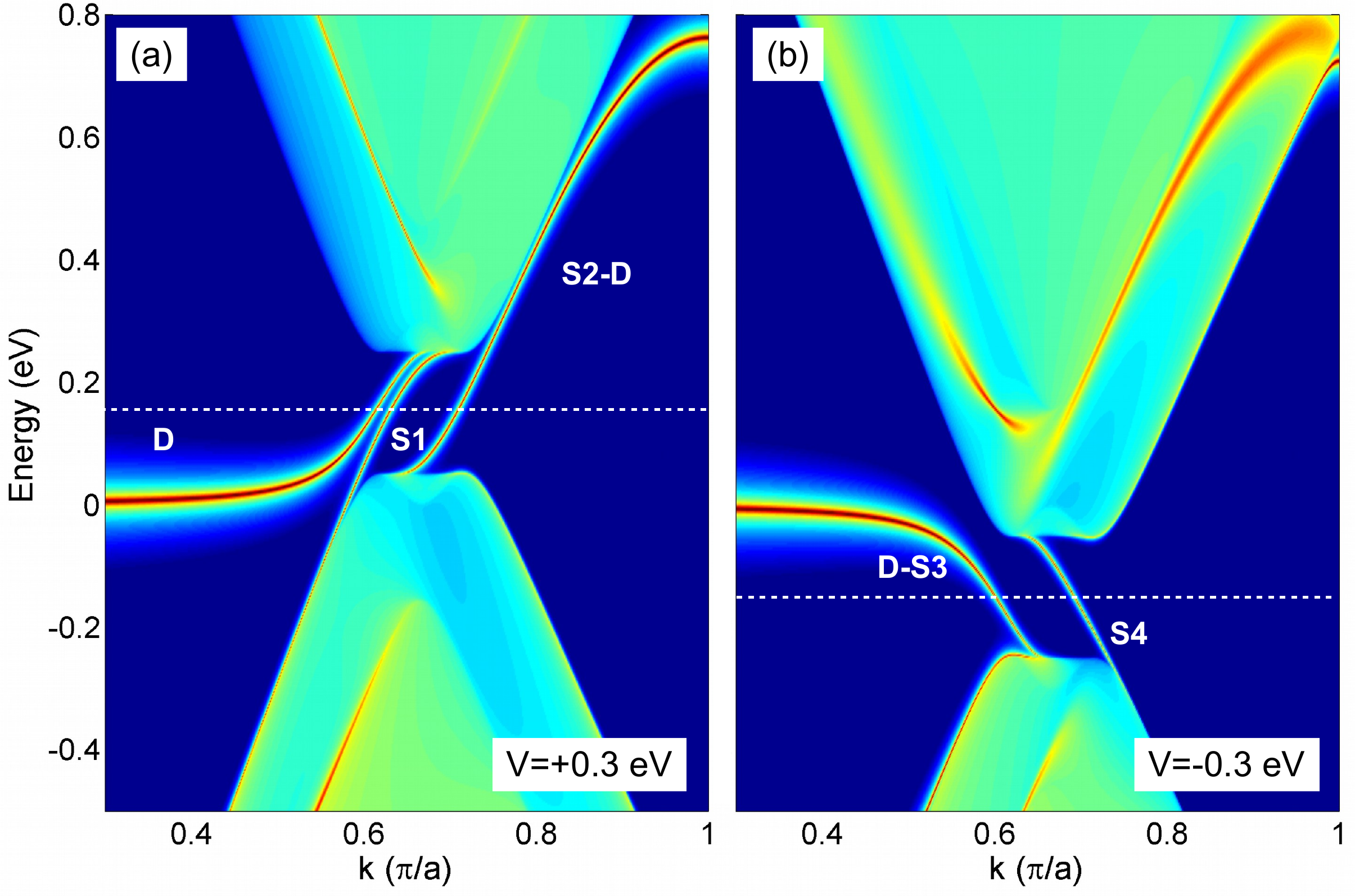}
\caption{\label{fig:ldosgbs}
(Color online). 
LDOS($E,k$) calculated at the defect line region of BLG. 
Gate voltage applied to the undefected layer is $V=0.3$ eV (a) and $V=-0.3$ eV (b). The wavevector $k$ corresponds to direction along the defect line in reciprocal units where $a$ is the unit cell width. Dashed lines mark the center of the gated bilayer gaps.
}
\end{figure}

\section{Ingredients}

On the one hand, previous works show that an 8-55 defect line in monolayer graphene yields a defect-localized band \cite{Lahiri_2010,Song,Liwei,Kan,Hu,our_prb_2013}.  In Fig. \ref{fig:ingr} (a) we show the LDOS as a function of energy and wavevector $k$ for an infinite monolayer graphene with this type of boundary. A defect flat band (D) spans from the Fermi level at $k=0$ to the Dirac point and next it emerges up to 0.8 eV at the edge of the BZ. The defect band to the left of the Dirac point is due to the zigzag edges facing each other, connected by the mixing nodes. These edge states appear at grain boundaries with topological defects, and we can relate them to the zigzag components of the joined edges \cite{ourprb2009,ourprb2011,our_prb_2013,Ayuela_2014}. Its localization changes from the Z-nodes of the pentagon (see Fig. \ref{fig:fig1}) to the mixing nodes at higher energies to the right of the Dirac cone. On the other hand, stacking boundaries have been shown to possess topological gap states when a gate voltage is applied throughout the system, if no-valley mixing symmetry holds. Fig. \ref{fig:ingr} (b) shows these topological bands when a positive (S1, S2) or a negative (S3, S4) voltage $V$ is applied to the bottom layer. These states can carry a current along the boundary which is valley-polarized. The sign of their velocities is changed by the reversal of the gate voltage \cite{Martin_2008}. In this work we investigate the interplay between the states produced by the octagon-double pentagon grain boundary and the topological gapless states resulting from the stacking change which, remarkably, persist after valley mixing.

\section{Results}
\label{sec:res}

In Fig. \ref{fig:ldosgbs} we present the LDOS($E,k$) at the boundary for energies around the bandgap. We focus on the valley with positive wavevectors near the Dirac point. For positive voltages, $V=0.3$ eV in panel (a), there are three bands in the energy gap. For negative voltages, $V=-0.3$ eV shown in panel (b), only two bands appear in the gap. The velocities of these bands change from positive to negative with the voltage sign. We can relate these states to the grain boundary band appearing in a single graphene layer and the two bands in bilayer graphene produced by the change of stacking (Fig. \ref{fig:ingr}). The flat band that is pinned at 0 eV for both voltages is clearly related to the defect line for low $k$. For positive $V$ (Fig. \ref{fig:ldosgbs} (a)) it follows closely another band, labeled S1, that has the same behavior as a topologically protected band in a simple stacking boundary, i.e., it joins the lower and the upper bulk continua. Similarly to the topologically protected band, the third band starts in the lower bulk continuum (S2), but crosses the gap without merging in the conduction band continuum, running parallel and close to it up to 0.8 eV, where it ends. Comparison of Fig. \ref{fig:ldosgbs} (a) and Fig. \ref{fig:ingr} allows us to infer that this band results from hybridization of the D and S2 bands, so we label it as S2-D. 

For $V<0$ the flat band at 0 eV, labeled D-S3, bends towards the lower bulk continuum, while the other band (S4) joins the upper and lower continua as a topologically protected band. Comparing Figs. \ref{fig:ingr} and \ref{fig:ldosgbs} (b), we infer that there are strong hybridization and mixing between the defect band D and the S3 band arising from the stacking change. Notice that despite this hybridization, we can still identify the parts of the band with D character. The band is again pinned at 0 V for low $k$ and fixed at 0.8 eV at the edge of the BZ, although it is now embedded in the continuum, which is shifted down by $V=-0.3$ eV.

\begin{figure}[thpb]
\centering
\includegraphics[width=\columnwidth]{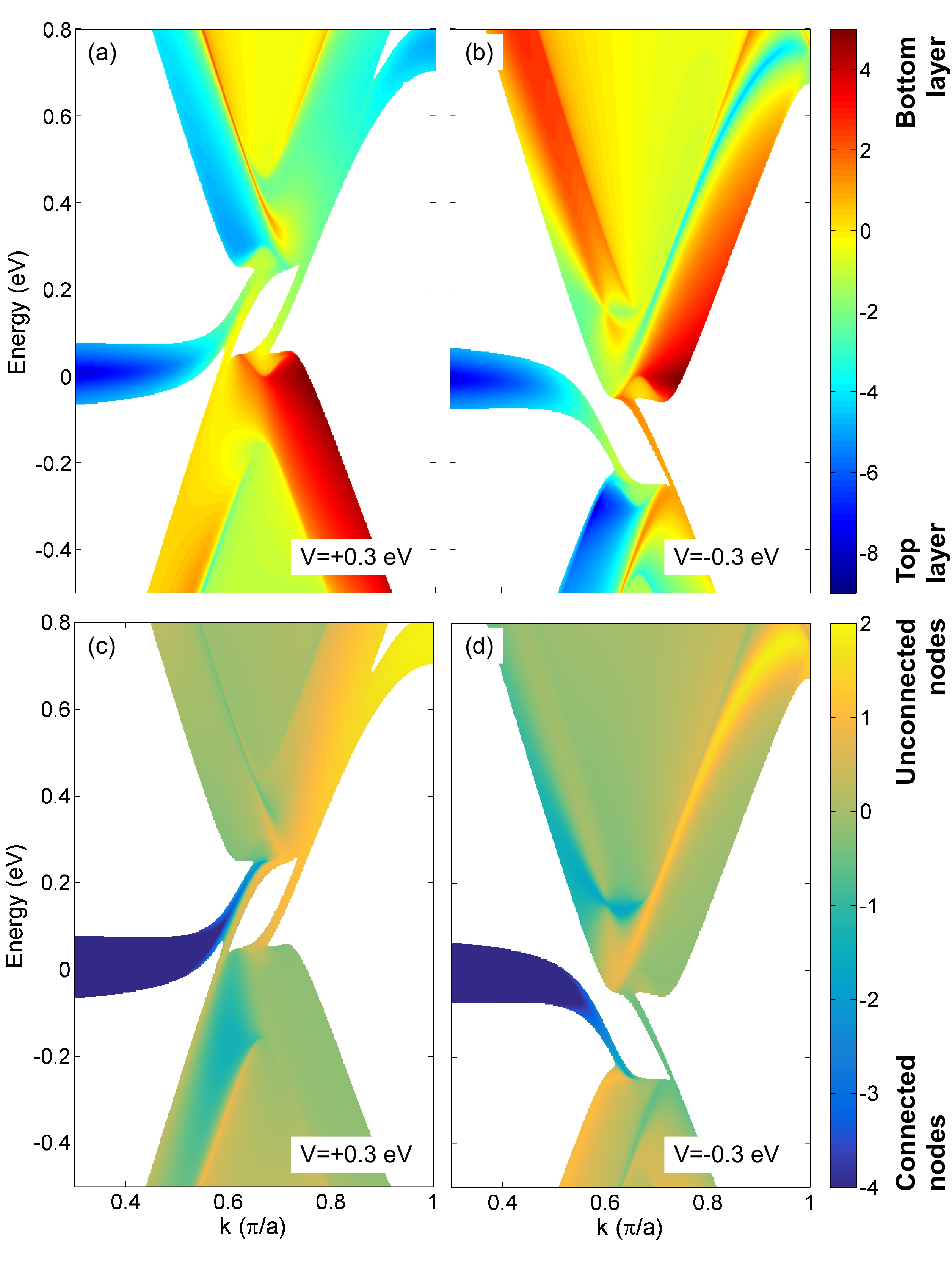}
\caption{\label{fig:figcolmap}
(Color online). 
(a) and (b): $\rm{ln}(\rm{LDOS}_B/\rm{LDOS}_T)$, where B/T refer to the bottom/top layers.
(c) and (d): $\rm{ln}(\rm{LDOS}_{U}/\rm{LDOS}_C)$ where U/C refer to unconnected/connected sublattices. The large width of the D-like portions of the bands at low and high $k$ is due to their strong localization.
}
\end{figure}

In order to understand the reasons why these different hybridizations take place, we present in Fig. \ref{fig:figcolmap} the LDOS resolved in top and bottom layers (top panels) and in connected/unconnected sublattices (lower panels). The defect band D that for $V>0$ joins the upper continuum at $k\approx \frac{2}{3}\frac{\pi}{a}$ has the same type of localization as the upper continuum, being both mostly in the top layer and in the connected sublattice. Likewise, for $k>\frac{2}{3}\frac{\pi}{a}$, the S2 band hybridizes with the D band that runs parallel to the conduction continuum, being both more localized in the top layer and in the unconnected sublattices. 

When the voltage is negative, the top/bottom localization of the two bulk continua is reversed, and the hybridization also changes. The D band, localized at the top layer, now strongly anticrosses with the S3 band, to the extent that we end up with a D-S3 band, which spans from a localized D to stacking S3 character, ending at the lower bulk continuum. As mentioned above, the higher $k$ portion of the defect band is buried in the upper continuum, barely emerging at 0.8 eV. The other band related to the change of stacking, S4, extends from the upper to the lower continuum, closing the gap. 

We verify that bands with similar localization, mostly at the same layer and preferentially in the same (connected or unconnected) sublattices, interact more strongly and hybridize, yielding in both cases bands that energetically close the gap, but they do not necessarily end in two continua. Instead, bands with a mixed novel character arise, that are D-like (defect-like) for some $k$ values and evolve into S-like (related to the stacking change) due to strong hybridization. It is important to note that for each sign of $V$, one of the bands originating from the stacking change (S1 or S4) preserves its strong and genuine topological character. 

%% FIGURE %%%

%% FIGURE %%%
\begin{figure}[thpb]
\centering
\includegraphics[width=\columnwidth]{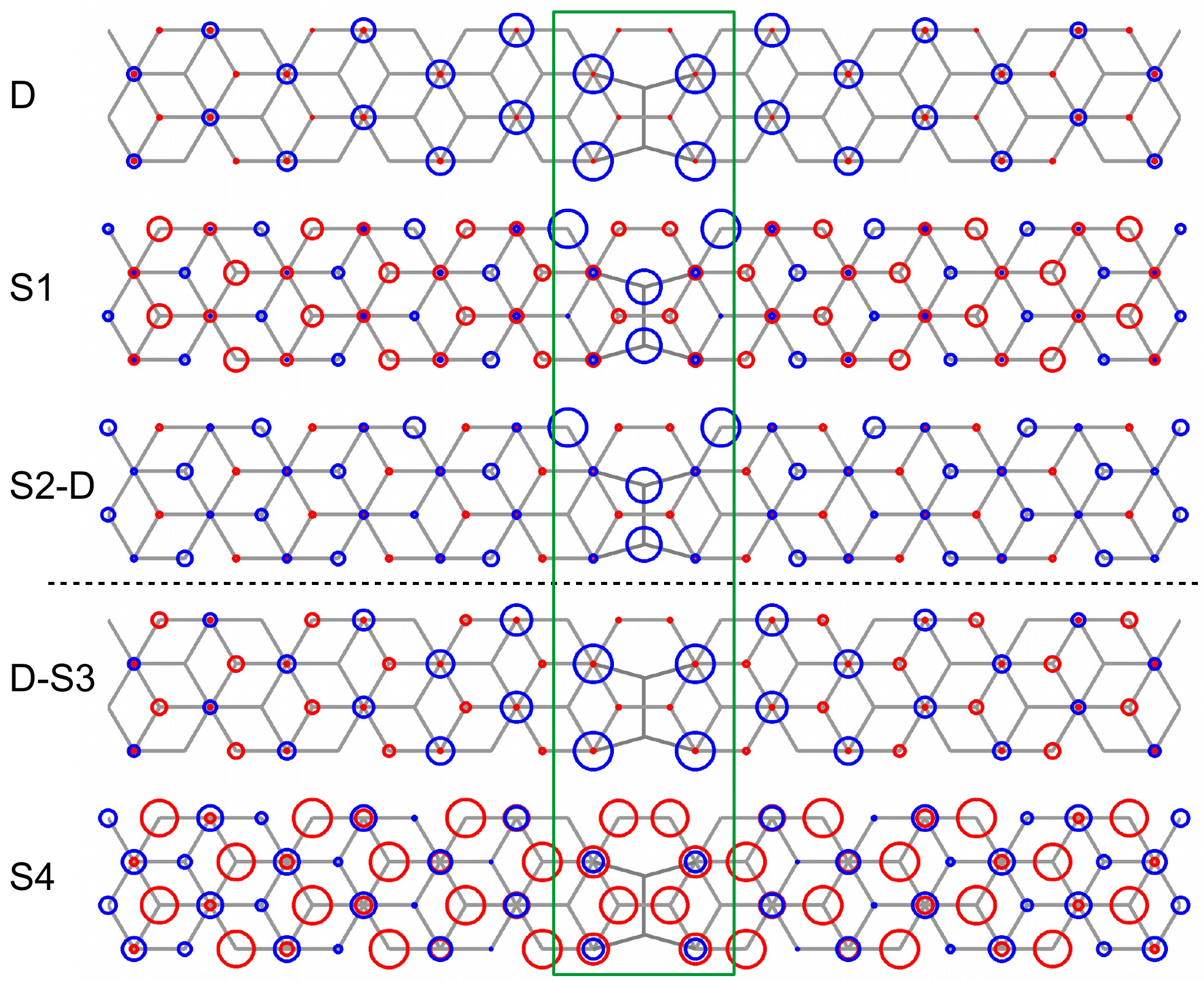}
\caption{\label{fig:fig4}
(Color online). Atom-resolved LDOS for the D, S1, S2-D, D-S3 and S4 gap states at the grain boundary region and surroundings. The LDOS is calculated for energies at the center of the gaps. The radii of the circles at the bottom (red) and top (blue) layers are proportional to the LDOS values. The green rectangle marks the grain boundary region with the 8-55 defects.
}
\end{figure}

We further analyze the role of the sublattices by resorting to the atom-resolved LDOS, plotted in Fig. \ref{fig:fig4}. The energy chosen corresponds to  the middle of the gap for both cases, marked with a dotted line in Fig. \ref{fig:ldosgbs}. For positive $V$ the D state is mostly localized at the connected sublattices, to which the Z-nodes belong. Although the S1 state has important weight at the top layer, as the defect state, it is mainly located at the unconnected sublattices, so they do not hybridize and run closely in parallel towards the upper continuum. The sublattice and layer distribution of S1 and S2-D are rather similar, and they share a remarkable feature: both are mostly localized at the mixing atoms. It is important to emphasize that this type of localization at low energies is not possible in a monolayer grain boundary \cite{our_prb_2013}; here it arises because of the AB/BA bilayer stacking change. It results in gap modes living at the unconnected sublattices, which are frustrated at the nodes responsible for the sublattice mixing.

For negative $V$, the D-S3 state is mostly located at the connected sublattices, with a predominant weight at the top layer, especially close to the topological defects. . We have checked that the unperturbed states D and S3 shown in Fig. \ref{fig:ingr} possess similar characteristics. 
 This explains the strong hybridization between the D and S3 bands, that leads to a band of mixed D-S3 character, as indicated in Fig. \ref{fig:ldosgbs}. The remaining state, S4, which preserves its topological character, has a larger LDOS in the bottom layer, with both sublattices having an appreciable weight.

If we focus on bilayer regions outside the green rectangle marking the grain boundary in Fig. \ref{fig:fig4}, the LDOS of the D, S2-D and D-S3 states 
is significantly concentrated in the upper layer, and shifts from the connected to unconnected nodes following the order D, D-S3, and S2-D as the component  
of the defect state (that is, degree of hybridization) decreases \cite{oscil}. In contrast, S1 and S4 topological states have a significant LDOS in the unconnected nodes of the bottom layer. Apparently, the S4 state, and to some extent also the S1 state, depicted in Fig. \ref{fig:fig4}, do not seem to be localized close to the grain boundary. This is an artifact due to the length of the plotted region. In fact, at longer distances than those shown in Fig. \ref{fig:fig4} they decay with slow oscillations. This is an interesting feature of these states appearing in the bulk gap and it is related to the fact that the decay is oscillatory because the Dirac point is not at $k=0$ \cite{Pelc_2015}. 

Similarly to AB/BA stacking boundaries produced by strain or a fold, the system presents a conductance gap measured perpendicularly to the grain boundary, but is conducting along it \cite{Ju_2015,Pelc_2015}. This longitudinal current is due to the gap states, and it is valley-polarized. If a local probe is used, such that only states along the defect line are measured, a conductance proportional to the number of states crossing the gap should be expected. Therefore, in contrast to previously studied boundaries \cite{Lin_2013,Alden_2013,Ju_2015, Pelc_2015}, here we have an asymmetry due to the different number of channels for the two gate polarizations, yielding a substantial change of one quantum of conductance (2$e^2 /h$) on the transport properties under gate reversal, that we have numerically verified. We propose that this distinctive asymmetric feature in the conductance of grain boundaries with 8-55 defects in BLG could be experimentally checked in a setup similar to the one presented in Ref. \cite{Ju_2015}. Furthermore, this characteristic could be also of interest for applications such as electrical switches, where the transport properties could be easily controlled by the sign of the applied gate voltage. Although the defect size is on the atomic scale in the direction perpendicular to the domain wall, in the direction of the boundary the defect is extended and periodic. So along this line the two valleys are well separated in reciprocal space and propagating states could travel without 
intervalley backscattering.
 
%%%%%%%%%%%%%%%%%%%%%%%%%%%%%%%%%%%%%%%%%%%%%%%%%%%%%%%%%%%%%%%%%%%%%%%%%%%%%%%%
%%% SUMMARY %%%
%%%%%%%%%%%%%%%%%%%%%%%%%%%%%%%%%%%%%%%%%%%%%%%%%%%%%%%%%%%%%%%%%%%%%%%%%%%%%%%%
\section{Conclusions}

In summary, we have shown that an AB/BA stacking boundary in bilayer graphene can be achieved by including a line of octagon and double-pentagon defects in one of the layers. Under an applied voltage a gap opens in the system and several robust gap states appear, unexpectedly, despite the presence of atomic-scale defects. We have elucidated the origin of these states, showing that in spite of the sublattice mixing due to the defects, one of the states preserves its full topological character for each gate polarization. It means that   topologically protected states can be much more common in bilayer graphene than so far predicted, since they can also appear at unexpected geometries, {\it e.g.}, at grain boundaries with atomic-scale  defects. The other gap states stem from the mixture of the band localized at the defect line with those originating from the stacking change. All of them provide conducting channels for the current flowing parallel to the defect line. More importantly, unlike previous predictions for stacking domain walls preserving the sublattice order, here the number of the gap states depends on the gate polarization. The change of the number of channels under gate polarization reversal yields a difference of conductance along the grain boundary that should be experimentally detected and could be exploited in electrical switches.


\begin{acknowledgements}
This work was partially supported by the Polish Ministry of Science and Higher Education (The "WZROST" Program) and the Spanish Ministry of Economy and Competitiveness under Grants No. FIS2012-33521, FIS2015-64654-P and FIS2013-48286-C2-1-P. We thank Dr. Garnett Bryant (National Institute of Standards and Technology,  Gaithersburg MD, and The University of Maryland) for critical reading of the manuscript and many valuable comments. 

\end{acknowledgements}

%\bibliography{bibliostacking}

\begin{thebibliography}{33}
\bibitem [{mis()}]{misc:LDOS}%
  \BibitemOpen
  \href@noop {} {}\bibinfo {note} {The LDOS is calculated in the region defined
  by the defect line in the top layer and the corresponding part of the bottom
  layer.}\BibitemShut {Stop}%
\end{thebibliography}

%merlin.mbs 2010-03-15 4.21a (PWD, AO, DPC)

\end{document}